\documentstyle[11pt,paspconf,epsf]{article}

\begin{document}

\title{Multiwavelength Studies of Interstellar Dust Using WIRE and MIGA}
\author{C. R. Kerton and P. G. Martin}
\affil{Department of Astronomy and CITA, University of Toronto, Toronto, ON, M5S 3H8, Canada}

\begin{abstract}
Studies of interstellar dust using the recently completed Mid-Infrared Galaxy Atlas (MIGA) and the soon to be launched Wide-Field Infrared Explorer (WIRE) satellite are summarized. Projects include looking at the distribution of dust around HII regions, the study of dust at the interfaces of HII regions and molecular clouds, and understanding the colour variations seen in cirrus clouds. 
\end{abstract}

\keywords{dust --- infrared radiation --- surveys --- HII regions}

\section{The Mid-Infrared Galaxy Atlas (MIGA)}
The Mid-Infrared Galaxy Atlas (MIGA) is a mid-infrared (12 and 25 $\mu$m) counterpart to the far-infrared Infrared Galaxy Atlas (IGA) (Cao et al.\ 1997). It consists of $1^{st}$ and $20^{th}$ iteration HIRES images, along with ancillary maps showing beam shape, coverage, photometric noise and IRAS detector tracks.

The basic steps involved in the construction of the MIGA are identical to those used in the construction of the IGA at 60 and 100 $\mu$m.  Preprocessing steps are done on a Sparc Ultra and YORIC is executed on a SGI Origin 2000.  As with the IGA, the large angular scale preprocessing of the IRAS data allows large high-quality mosaics to be constructed from the final HIRES images.  The MIGA and IGA will be the infrared data sets for the Canadian Galactic Plane Survey (CGPS), a project to survey about a quadrant ($75^{\circ} < l < 148^{\circ}$) of the Galactic Plane at about $1'$ resolution over a wide range of wavelengths and ISM componets. (e.g., English et al.\ 1998).

To date, a region corresponding to the CGPS survey region in Galactic longitude and  $\pm6^{\circ}$ in latitude has been processed.  Far-infrared images in the CGPS region above the $+4.7^{\circ}$ limit of the IGA have also been  produced to provide complete IR coverage of the CGPS region.  Supplementary processing in the immediate future is focussing on select fields in support of the WIRE mission and other projects being undertaken by the CGPS consortium.

\section{The Wide Field Infrared Explorer (WIRE)}
WIRE (Wide-Field Infrared Explorer) is a small (30 cm aperture, f/3.3, Ritchey-Chretien) infrared astronomical satellite developed as part of NASA's Small Explorer program. The telescope will be able to simultaneously image a $33'\times 33'$ field of view in two colour bands centered at 12 and 25 $\mu$m (9--15 and 21--27$\mu$m respectively). Spatial resolution will be $\sim20''$ at 12$\mu$m and $\sim23''$ at 25$\mu$m.  For detailed information on the spacecraft and instruments see:

\begin{center}
http://www.ipac.caltech.edu/wire/.
\end{center}

\begin{figure}
\plotone{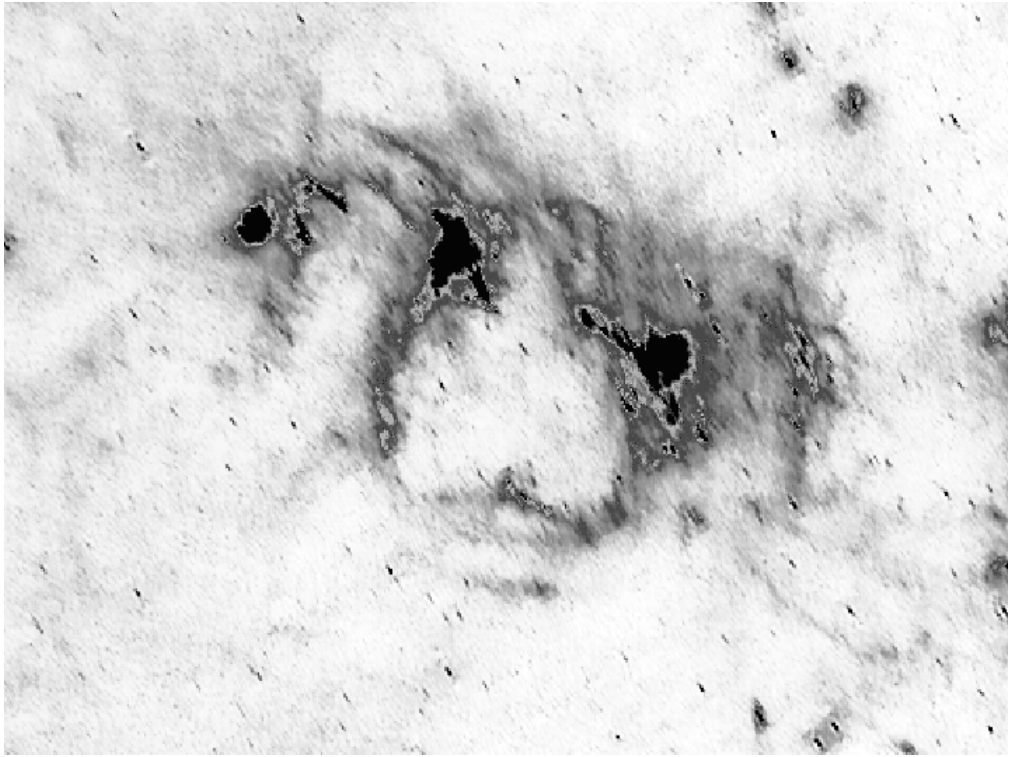}
\caption{The W5 HII Region at 12$\mu$m. (see \S 4)}
\end{figure}

Although the primary purpose of the mission is the study of galaxy evolution, part of the observing time has been made available through an Associate Investigator (AI) program in order to study targets near the Galactic Plane and the Ecliptic, areas where WIRE cannot carry out its primary science survey. We are co-investigators (along with G. Helou, C. Beichman, R. Cutri and M. Moshir at IPAC, and A. Abergel, J.-Ph. Bernard and F. Boulanger at IAS) on an AI proposal to study the evolution of dust within the ISM.

Currently WIRE is scheduled for launch in March 1999 from Vandenberg AFB for a four month mission.

\section{Science With WIRE and MIGA}
The two mid-infrared wavelength bands covered by MIGA (IRAS) and WIRE are well placed to study the evolution of dust in the ISM. It is now well established that the 12$\mu$m diffuse emission measured by IRAS is dominated by emission bands considered to be characteristic of PAHs.  Spectra obtained by IRTS and ISO show the presence of the full set of PAH features for a wide range of clouds (Onaka et al.\ 1996, Mattila et al.\ 1996). The emission at 25$\mu$m comes from particles intermediate in size between PAHs and classical dust grains. Since the chemical nature of these particles is as yet unknown, they are simply referred to as Very Small Grains or VSGs. 

WIRE will be able to map the diffuse emission at 12 and 25 $\mu$m for a wide range of physical conditions with high resolution and sensitivity (5$\sigma = 0.2$ MJy sr$^{-1}$ at full resolution).  The 25$\mu$m images should be especially interesting since this band was not available on ISO for imaging. WIRE's wide field of view will also allow large (few degrees in scale) mosaics to be constructed.

MIGA, in spite of its lower sensitivity and the artifacts associated with the HIRES process, will be useful for looking at features that WIRE cannot view due to their brightness or because of mission constraints (e.g., lack of suitable guide stars). At best, we will be able to look at only about 25 different regions using WIRE. MIGA will be especially useful for looking at bright HII regions.

WIRE and MIGA will be used to look at the evolution of dust within dense, translucent and diffuse clouds as well as dust associated with HII regions.  By comparing maps made at 12 and 25 $\mu$m we can study how PAH and VSG emission depend on local physical conditions.  Colour variations at these wavelengths will provide insight into both the local heating processes and the relative abundance of PAHs and VSGs in the ISM (Boulanger et al.\ 1990).  Complementary radio and mm data will be used to characterize different physical environments and radiation fields accurately.
 
We are just starting to use MIGA to look at Sharpless HII regions within the CGPS survey area.  The initial sample has been limited to those regions associated with a single exciting star and $< 60'$ in angular diameter.  These criteria were chosen to simplify modelling of the HII region's energetics and to minimize the computational time involved in creating HIRES colour maps.  We plan to look at larger regions and areas outside of the CGPS survey area in the future. For the latest information on this project see:

\begin{center}
http://www.utoronto.cita/$\sim$kerton
\end{center}

\section{A Sample MIGA Region - W5}
W5 is a star-forming region located in the Perseus Arm of the Galaxy at $l=137.5^{\circ}, b=1.0^{\circ}$.  It is made up of two regions: W5-East, excited by a single O star, and W5-West, excited by a cluster of four O stars.  The view shown in Figure 1 is a $4^{\circ}\times3^{\circ}$ mosaic of $20^{th}$ iteration HIRES images at 12$\mu$m. The image shows the enhanced infrared emission occurring just beyond ionization fronts in both regions.  Images at 25$\mu$m show that the emission is much more extended, and a detailed comparison of the morphology is currently underway.

\section{A Sample WIRE Region - Octopus}
An example of an interesting class of WIRE targets, clouds that have strong infrared colour variations, is shown in Figure 2.  Determining the cause of these colour variations is an important goal of our WIRE proposal. The WIRE observations for Octopus are centered at $l = 346.8$ $b = 20.9$.  An added bonus is that these observations will include smaller areas that were also imaged with ISO.

\begin{figure}
\plotone{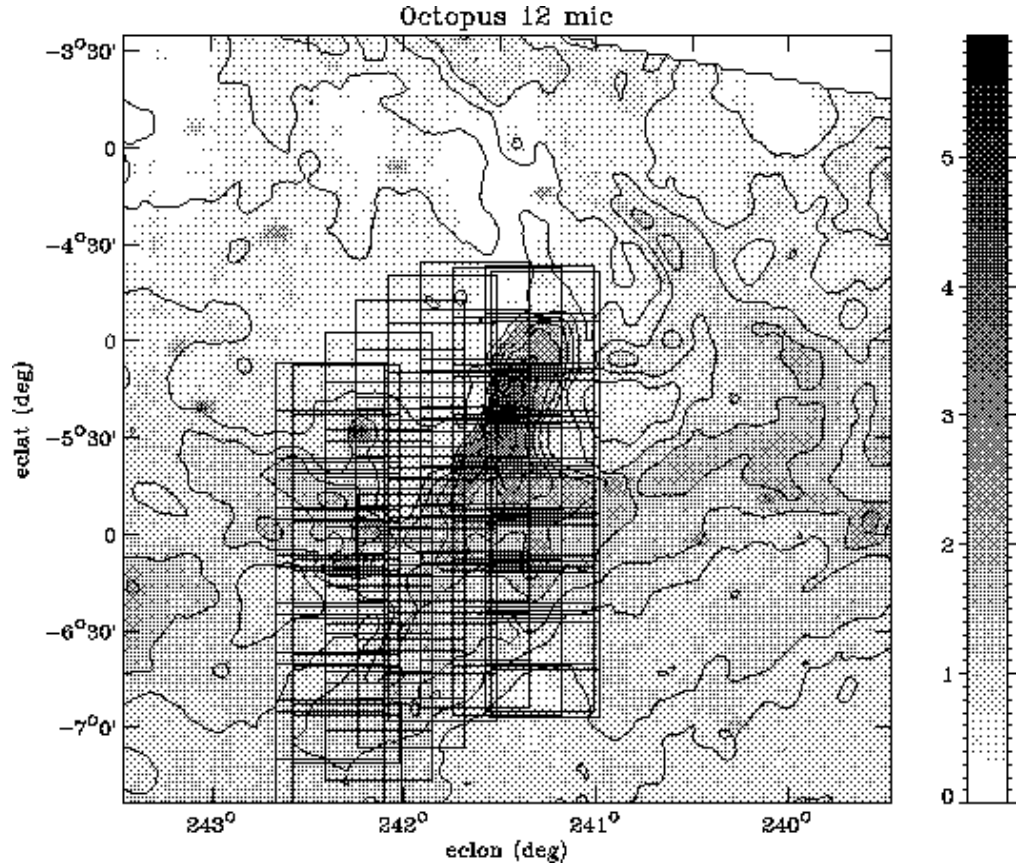}
\caption{Octopus, one of the clouds with strong infrared colour variations to be looked at with WIRE (see \S 5). Mid-infrared emission at 12$\mu$m is shown in greyscale, with contours of 100$\mu$m emission (both from ISSA). The areas to be observed by WIRE are indicated by the overlapping boxes.  Units are MJy sr$^{-1}$}
\end{figure}

\end{document}